\documentstyle[seceq]{ptptex}

\title{
The light-cone gauge without prescriptions
}

\author{
Alfredo Takashi {\sc Suzuki}\footnote{E-mail:suzuki@ift.unesp.br} 
and A.G.M.{\sc Schmidt}\footnote{E-mail:schmidt@ift.unesp.br}
}

\inst{
Instituto de F\'{\i}sica Te\'orica, Universidade Estadual Paulista,
R.Pamplona, 145 \\ S\~ao Paulo - SP CEP 01405-900 Brazil}

\recdate{
}

\abst{
Feynman integrals in the physical light-cone gauge are harder to  solve than
their covariant counterparts. The difficulty is associated with the presence of
unphysical singularities due to the inherent residual gauge freedom in the
intermediate boson propagators constrained within this gauge choice. In order
to circumvent these non-physical singularities, the headlong approach has
always been to call for mathematical devices --- prescriptions --- some
successful ones and others not so much so. A more elegant approach is to
consider the propagator from its physical point of view, that is, an object
obeying basic principles such as causality. Once this fact is realized and
carefully taken into account, the crutch of prescriptions can be avoided
altogether. An alternative third approach, which for practical computations
could dispense with prescriptions as well as  prescinding the necessity of
careful stepwise watching out of causality would be of great advantage. And
this third option is realizable within the context of negative dimensions, or
as it has been coined, negative dimensional integration method (NDIM).
}
\begin{document}

\maketitle

\section{Introduction}

Light-cone gauge in quantum field theory is a fascinating subject. It is a
``simple'' gauge choice in the sense that the emerging propagator for the boson
field has a deceivingly simple structure, when compared to other non-covariant
gauge choices. Yet, as it has soon been realized, it hid subtle complications
when one actually wanted to work with it.

To make things more concrete, let us analyse it in the framework of vector
gauge fields, e.g. the pure Yang-Mills fields, where, after taking the limit of
vanishing gauge parameter, the propagator reads,

\begin{equation}
\label{prop}
D^{ab}_{\mu\nu}(k)=\frac{-i\,\delta^{ab}}{k^2+i\varepsilon}
\left[g_{\mu\nu}-\frac{n_{\mu}k_{\nu}+n_{\nu}k_{\mu}}{k\!\cdot\! n} \right ]
\end{equation}
where $n_{\mu}$ is the arbritary and constant null four-vector which defines
the gauge, $n\!\cdot\!A^a(x)=0;\;n^2=0$. This generates $D$-dimensional
Feynman integrals of the following generic form,

\begin{equation}
\label{struc}
I_{lc}=\int
\frac{d^Dk_i}{A(k_j,\;p_l)}\frac{f(k_j\!\cdot\!n^*,\;p_l\!\cdot\!n^*)}{h(k_j\!\cdot\!n,\;p_l\!\cdot\!n)},
\end{equation}
where $p_{l}$ labels all the external momenta, and $n^{*}_{\mu}$ is a null
four-vector, dual to $n_{\mu}$. We will evaluate, as a pedagogical example, an
integral where, 
\begin{equation} \label{exemplo} A(k_j,\;p_l) =
(k^2)^{-i}\left[(k-p)^2\right]^{-j}, \;\; h(k_j\!\cdot\!n,\;p_l\!\cdot\!n) =
(k\cdot n)^{-l},\;\; f(k_j\!\cdot\!n^*,\;p_l\!\cdot\!n^*) = (k\cdot n^*)^m, 
\end{equation} with $i,j,l$ negative and $m$ positive or zero.

A conspicuous feature that we need to note
first of all, is that the dual vector $n_{\mu}^{*}$, when it appears at all,
it does so {\em always } and {\em only} in the numerators of the integrands.
And herein comes the first seemingly ``misterious'' facet of light-cone gauge.
How come that from a propagator expression like (\ref{prop}), which contains no
$n^*$ factors, can arise integrals of the form (\ref{struc}), with prominently
seen $n^*$ factors? Again, this is most easily seen in the framework of 
definite external vectors $n$ and $n^*$. An alternative way of writing the
generic form of a light-cone integral is

\begin{equation}
\label{struc1}
I_{lc}^{\mu_1\cdots \mu_n}=\int
\frac{d^{D}k_{i}}{A(k_{j},\;p_{l})}\frac{g(k^{\mu_j}_{j},\;p_l^{\mu_l})}{h(k_{j}\!\cdot\! n,\;p_{l}\!\cdot\! n)},
\end{equation}
where the numerator $g(k^{\mu_j}_{j},\;p_l^{\mu_l})$ defines a tensorial
structure in the integral. For a vector, we have $k^{\mu}=(k^+,\;k^-,\;{\bf
k}^{\sc t})$, where $k^+=\lambda(k^0+k^{D-1})$ and  $k^-=\lambda(k^0-k^{D-1})$
with $\lambda$ being a normalization factor. If we choose definite $n$ and
$n^*$ such that $n_{\mu}=(1,\;0,\;\cdots,\;1)$, and
$n^*_{\mu}=(1,\;0,\;\cdots,\;-1)$, this allows us to write $k^+\equiv
k\!\cdot\!n$ and $k^-\equiv k\!\cdot\!n^*$. We have therefore traced back the
origin for the numerator factors containing $n^*$. By the way, these terms have
nothing whatsoever to do with some kind of prescription input as it is in
(\ref{presc}) below. There, the $n^*$ factors are added into the structure {\em
by hand} via the {\em ad hoc} prescription.

\section{Prescriptions}

Now, the troublesome factors in the denominator, represented by
$h(k_{j}\!\cdot\!n,\;p_{l}\!\cdot\! n)$, for one-loop, two-point functions are
typically products of the form

\begin{equation}
\frac{1}{h(k_{j}\!\cdot\!n,\;p_{l}\!\cdot\! n)} \sim \frac {1}{(k\cdot n)\:(k-p)\cdot n}.
\end{equation}

In the standard approach, these are dealt with first by partial fractioning
them, a trick sometimes called ``decomposition formula'' (see, for example,
\cite{leib,leib-nyeo,leib-nyeo2})

\begin{equation}
\label{decomp}
\frac {1}{(k\!\cdot\! n)\:(p-k)\!\cdot\! n}\rightarrow \frac{1}{p\!\cdot\!
n}\left[\frac{1}{k\!\cdot\! n}+\frac{1}{(p-k)\!\cdot\! n}\right ]\:,
\:\:\:\:\:\:\:p\!\cdot\! n \neq 0
\end{equation}

It is easy to see that such kind of trick becomes very clumsy to deal with when
one has higher powers of denominator factors. In fact, the authors of
\cite{leib-nyeo2} agree that the application of (\ref{decomp}) complicates
considerably the calculation of such Feynman integrals. For example, consider,

\begin{equation}
\label{decomp1}
\frac {1}{(k\!\cdot\! n)^2\:(p-k)\!\cdot\! n}\rightarrow \frac{1}{(p\!\cdot\!
n)^2}\left[\frac{p\!\cdot\!n}{(k\!\cdot\!n)^2}+\frac{1}{k\!\cdot\! n}+\frac{1}{(p-k)\!\cdot\! n}\right ]\:,
\:\:\:\:\:\:\:p\!\cdot\! n \neq 0
\end{equation}
and so on and so forth. Yet, the standard approach based upon prescriptions,
cannot handle denominator products like these without first decomposing them.
As far as we know only Leibbrandt and Nyeo said explicitly before,
that ``decomposition formulas'' like (\ref{decomp}) and (\ref{decomp1}) are in
fact part of the prescription\cite{leib-nyeo2}. Be it as it is, after the
decomposition is made, {\em then and only then}, one seeks to handle the
isolated denominators calling for prescriptions that would turn them
manageable mathematically.

Early efforts in this direction draw heavily from the use of the Cauchy
principal value  (CPV or PV for short),  
\begin{equation} 
PV \left(\frac{1}{q\!\cdot\! n}\right) \longrightarrow \frac{1}{2} \lim_{\epsilon\to 0} \left(  
\frac{1}{q\!\cdot\! n +i\epsilon}+ \frac{1}{q\!\cdot\! n - i\epsilon} \right) . 
\end{equation} 
Unfortunately, this prescription, although preserves the general structure of
the light-cone integral (\ref{struc}), and is mathematically consistent, does 
not provide physically acceptable results in the light-cone
gauge\cite{bass,bass2}.   

Later on, independently Mandelstam\cite{mandelstam} and Leibbrandt\cite{leib2} 
proposed new prescriptions to treat the gauge-dependent poles,
  
\begin{eqnarray} 
\label{presc}
\frac{1}{q\!\cdot\! n} &\longrightarrow &\frac{1}{q\!\cdot\! n +i\epsilon \;
sgn (q\!\cdot\! n^*)}\nonumber \\
& = & \frac{q\!\cdot\! n^*}{(q\!\cdot \! n) \, (q\!\cdot \! n^*) +i\epsilon}. 
\end{eqnarray} 
where the former is the modified Mandelstam prescription and the latter is the
Leibbrandt's one. It can be proven that, in fact, they  are completely
equivalent to each other, so that sometimes people refer to this as the
Mandelstam-Leibbrandt (ML) prescription. A conspicuous feature of these
prescriptions is that they introduce the factor $(q\cdot n^*)$ in the
denominator of integrals, thus modifying the original structure. This, however,
does not seem to affect the end result, as far as physically meaningful results
are concerned. 

Applying the ML prescription for example, in the one-loop two-point function
integrals, double poles still appear but they cancel against each other and one
is left with physical poles only. In other words, those singularities that are
merely gauge artifacts cancel out when one uses this approach.
Experts\cite{bass2} argue, therefore, that adopting the ML prescription makes
the light-cone gauge acceptable, at least perturbatively. Along this line, some
two-loop calculations have been performed explicitly and can be found in the
pertinent literature\cite{twoloop}.

Pimentel {\em et al.} realized that when the physical principle of causality is
correctly taken into account --- either by careful, stepwise watching out for
it to be preserved along the whole computation, or by implementing it directly
onto the propagator, via causal considerations the same picture arises, namely,
unphysical, gauge-dependent poles cancel out leaving solely physical
ones\cite{pimentel,covariantization}. Actually, the procedure corresponds
exactly to taking out the zero-frequency mode from the Fourier series expansion
for the field operators, ensuring that only positive energy quanta propagate into
the future and vice versa. Mathematically this can be acomplished via

\begin{equation}
\frac{1}{(q\!\cdot\! n)^j} = PV\left(\frac{1}{(q\!\cdot\! n)^j}\right) -i\pi 
\frac{(-1)^{j-1}}{(j-1)!} \delta^{(j-1)}(q\!\cdot\! n) \;sgn(q^0) , 
\end{equation} 
where $PV$ stands for Cauchy principal value. The second term of this expression
is crucial. It is the term that guarantees that the troublesome zero-frequency
mode of the field quanta is subtracted out from the energy spectrum. It is
exactly tantamount to avoiding the pathological mixing of quanta of opposite
energy signs propagating into the future, thus violating causality.

By the very nature in which these prescriptions are introduced, they are
completely powerless to deal with products of light-cone poles without ere
using the partial fractioning trick. Moreover, in addition to the need to use
the ``decomposition formula'', the necessity to resort to prescriptions or even
clever maneuvers in order to extract out the culprit factor from the original
integral is an awkward trail to walk along to reach our objective. Because a
prescription is solely a prescription, a mere device to by-pass a problem,
prescriptions are usually inconvenient and undesirable. Yet, as awkward and
burdensome as it might be, there was no other way we could do things
properly in dealing with perturbative light-cone gauge computations.

\section{Negative dimensional approach}

Not so now, with the advent of NDIM\cite{halliday}. NDIM is a technique wherein
the principle of analytic continuation plays a key role. With it we solve a
``Feynman-like'' polynomial fermionic integral, i.e., a loop integral in
negative $D$-dimensional space with propagators raised to positive powers in
the numerator. Solutions arise as linear combinations of solutions for simple
systems of linear equations and these are then analytically continued to allow
for negative values of exponents (i.e. propagators now become raised to
negative powers, becoming denominator terms in the integrands) and positive
dimension\cite{flying}. This procedure, of course, when applied to the
light-cone case has to consider the general structure (\ref{struc}) above which
is characteristic of this gauge. That is, terms like $f(k_{j}\!\cdot\!
n^{*},\;p_{l}\!\cdot\! n^{*}) \sim (q\!\cdot\! n^*)^a[(q-p)\!\cdot\! n^*]^b$
which can appear in the numerator, evidently must remain there; therefore, no
exponents of such terms are to be analytically continued to allow for negative
values\cite{probing}. Schematically (see for instance second reference in 
\cite{flying}),

\begin{equation} 
\label{aci}
\int d^{D}k_{i}\;A(k_{j},\;p_{l})\;f(k_{j}\!\cdot\!
n^{*},\;p_{l}\!\cdot\! n^{*})\:h(k_{j}\!\cdot\! n,\;p_{l}\!\cdot\! n)\longrightarrow 
^{\!\!\!\!\!\!\!\!\!\!\!\!AC}\int
\frac{d^{D}k_{i}}{A(k_{j},\;p_{l})}\frac {f(k_{j}\!\cdot\!
n^{*},\;p_{l}\!\cdot\! n^{*})}{\:h(k_{j}\!\cdot\!
n,\;p_{l}\!\cdot\! n)}
\end{equation} 
where the left-hand side shows the negative dimensional integral to be
performed and the right-hand side displays the generic form (\ref{struc}) for
the light-cone integrals. Note that the factor $f(k_{j}\!\cdot\!
n^{*},\;p_{l}\!\cdot\! n^{*})$ remains in the numerator of the integrands in
both sides. The left-hand side of (\ref{aci}), which is evaluated via NDIM
methodology, is defined from projecting out powers of Gaussian type
$D$-dimensional momentum integrals of propagators\cite{halliday}. It is worth
mentioning here that the usual Schwinger exponentiation of propagators for the
factor $[h(k_{j}\!\cdot\! n,\;p_{l}\!\cdot\! n)]^{-1}$ in positive dimensional
calculation does not work for light-cone integrals \cite{covariantization}.

Now, one could rightfully ask: how can one get the standard ML results for one
and two loop ML dimensionally regulated Feynman-integrals with the procedure of
negative dimensions? Substitute (\ref{exemplo}) in (\ref{struc}), that is,
consider the integral,

\begin{equation} 
B(i,j,l,m) = \int d^D\!q\;
(q^2)^i\left[(q-p)^2\right]^j(q\cdot n)^l (q\cdot n^*)^m. 
\end{equation}
Let us evaluate it in great detail following the steps thoroughly described in
our previous papers\cite{flying}. Let our starting point be, 
\begin{eqnarray} 
\label{Gaussian}
G &=& \int
d^D\!q\; \exp{\left[ -\alpha q^2 - \beta (q-p)^2 -\gamma q\cdot n -\theta
q\cdot n^*\right]}\nonumber\\ && = \left(\frac{\pi}{\lambda}\right)^{D/2}
\exp{\left[-\frac{1}{\lambda} \left( \alpha\beta p^2 + \beta\gamma p\cdot n +
\beta\theta p\cdot n^* -\frac{1}{2}\gamma\theta n\cdot
n^*\right)\right]},
\end{eqnarray} 
where $\lambda=\alpha+\beta$. Taylor
expanding the exponentials above and solving for the integral $B(i,j,l,m)$ we
obtain, 
\begin{eqnarray} B(i,j,l,m) &=& (-\pi)^{D/2} i!j!l!m!(-\sigma-D/2)!
\sum_{\{X,Y\}=0}^\infty
\frac{(p^2)^{X_1}(p^+)^{X_2} (p^-)^{X_3}}{X_1!X_2!X_3!X_4!Y_1!Y_2!}
\left(\frac{-nn^*}{2}\right)^{X_4}\nonumber\\
&&\times \delta_{a,i}\delta_{b,j}\delta_{c,l}
\delta_{d,m}\delta_{e,\sigma} ,\end{eqnarray}
where $\sigma=i+j+l+m+D/2$, $a=X_1+Y_1$, $b=X_1+X_2+X_3+Y_2$, $c=X_2+X_4$,
$d=X_3+X_4$, $e=X_1+X_2+X_3+X_4$. This system of linear algebraic equations
has six possible solutions -- since the number of  unknowns is bigger than the
number of equations. In all six cases eliminating the deltas leaves us with
one remaining sum, a hypergeometric series $_3F_2$. Three out of these six
series have as its variable,
$$ z=\left(\frac{p^2 nn^*}{2p^+ p^-}\right),$$
while the other three have as variable $$
w \equiv z^{-1}=\left(\frac{2p^+ p^-}{p^2 nn^*}\right) .$$

In our previous work \cite{probing}, we have considered the simplest one-loop
integral in the light-cone gauge, evaluating it via NDIM method, with one- and
two-degree violation of Lorentz covariance (more specifically, with $n_\mu$
only and with both $n_\mu$ and its dual $n_\mu^{*}$). For the first case we
obtained the usual PV prescription result, whereas in the second case we
obtained the Mandelstam-Leibbrandt prescription result. However, that toy
integral is easily evaluated in the NDIM scheme, since there are no left over
summation indices, that is, the number of linear equations in the system equals
the number of unknowns and the result can be expressed just in terms of product
of gamma functions. The integrals we consider here are more complicated; they
generate several linearly independent and dependent solutions which need to be
sorted out carefully. That is, the space of solutions spanned by base functions
are not all linearly independent ones. This is characteristic of these
particular cases of integrals we are considering here, and this is the reason
we treat them at lenght and in depth.    

From our previous works \cite{flying} we know that the Feynman integral will
be represented by a linear combination of linearly independent series.
Moreover, these representations are related through analytic continuation. Let
us consider the representation of $B(i,j,l,m)$ in terms of $z^{-1}$,
\begin{equation} B(i,j,l,m) = B_1 + B_2 ,\end{equation}
where
\begin{eqnarray} B_1 &=& (-\pi)^{D/2}\frac{\Gamma(1+j)\Gamma(1+l)
\Gamma(1+m)(p^2)^i(p^+)^{j+l+D/2} (p^-)^{j+m+D/2}}{\Gamma(1+j+l+D/2)
\Gamma(1+j+m+D/2) \Gamma(1-j-D/2)}\nonumber\\ 
&&\times\left(\frac{-nn^*}{2}\right)^{-j-D/2} \,
_3F_2(a_1,b_1,c_1;e_1,f_1|w) ,\end{eqnarray} 
and 
\begin{eqnarray} B_2 &=& (-\pi)^{D/2}\frac{\Gamma(1+i)\Gamma(1+j)
\Gamma(1+m)\Gamma(1-\sigma-D/2) (p^2)^{\sigma-m}
(p^-)^{-l+m}}{\Gamma(1-j-l-D/2) \Gamma(1-i-m-D/2)
\Gamma(1-l+m)\Gamma(1+\sigma-m)} \nonumber\\
&&\times\left(\frac{-nn^*}{2}\right)^{l}
\, _3F_2(a_2,b_2,c_2;e_2,f_2|w) ,\end{eqnarray}
where $a_1=-i$, $b_1=D/2+j$, $c_1=\sigma+D/2$, $e_1=1+j+l+D/2$,
$f_1=1+j+m+D/2$, $a_2=-l$, $b_2=-i-j-l-D/2$, $c_2=i+m+D/2$, $e_2=1-l+m$,
$f_2=1-j-l-D/2$ and $_3F_2$ is the generalized hypergeometric
function\cite{luke}. This is the result in the negative dimension region and
$i,j,l,m$ positive. Now, in order to be physically meaningful it must be
analytically continued to positive dimension and $i,j,l$ negative, while the
exponent $m$ must be kept untouched. To carry out this analytic continuation
one must rewrite the gamma factors as Pochhammer symbols and use one of its
properties  \begin{equation}
(a|j) \equiv (a)_j = \frac{\Gamma(a+j)}{\Gamma(a)}, \;\;\;\; (a|-j) =
\frac{(-1)^j}{(1-a|j)}. \end{equation}
Performing these simple operations one gets,
\begin{eqnarray}
B^{AC}(i,j,l,m) &=& \pi^{D/2}
\frac{(-j|-l-D/2)(-l|j+l+D/2)}{(1+m|j+D/2)} (p^2)^i (p^+)^{j+l+D/2}\nonumber\\
&&\times(p^-)^{j+m+D/2}\left(\frac{nn^*}{2}\right)^{-j-D/2}\,
_3F_2(a_1,b_1,c_1;e_1,f_1|w) \nonumber\\
 &&+\pi^{D/2}
\frac{(-j|i+j+m+D/2)(-i|i+j+l+D/2)}{(1+m|-l)}\nonumber\\
&&\times (\sigma+D/2| -2\sigma-D/2+m) (p^2)^{\sigma-m}
(p^-)^{-l+m} \left(\frac{nn^*}{2}\right)^{l}\nonumber\\
&&\times \,_3F_2(a_2,b_2,c_2;e_2,f_2|w). \end{eqnarray}

In fact, from the theory of hypergeometric functions\cite{luke}, we know that the hypergeometric differential equation for
$_pF_q$ has up to $p$ linearly independent solutions, so, in writing
down $B(i,j,l,m)$ we would expect it to contain up to three terms. Indeed,
among the three series of funtions $_3F_2(\cdots|w)$ there is a third
term, namely,
 \begin{eqnarray} B_3 &=&
(-\pi)^{D/2}\frac{\Gamma(1+i)\Gamma(1+j)
\Gamma(1+l)\Gamma(1-\sigma-D/2)(p^2)^{i+j+m+D/2}
}{\Gamma(1+i+j+m+D/2)\Gamma(1+l-m) \Gamma(1-j-m-D/2)
}\nonumber\\
&& \times \frac{(p^+)^{l-m}}{\Gamma(1-i-l-D/2)}
\left(\frac{-nn^*}{2}\right)^{m}\,_3F_2(a_3,b_3,c_3;e_3,f_3|w) ,
\end{eqnarray}
where $a_3=-i-j-m-D/2$, $b_3=-m$, $c_3=i+l+D/2$, $e_3=1+l-m$,
$f_3=1-j-m-D/2$. However, this solution is not a linearly independent one.
Indeed, from the fact that $m$ is always positive or zero we can
rewrite $B_3$ using eq.(6), section 3.2 of ref.\cite{luke} --- if we choose
$\alpha_1=c_3$, $\alpha_2=a_3$, $n=-e_3$, $\rho_1=f_3$ --- as a linear
combination of $B_1$ and $B_2$. 

The integral we are considering here was studied by Lee and Milgram\cite{lee}. Our result
can also be written in terms of the so-called $G-$function or Meijer's
function and agrees with their calculation. Taking the special case where $i=j=l=-1$
and $m=0$ one gets,
\begin{eqnarray}
B(-1,-1,-1,0) &=& -(-\pi)^{D/2}\frac{\Gamma(2-D/2)\Gamma(D/2-1)}{\Gamma(D/2)}
\frac{(p^+)^{D/2-2}(p^-)^{D/2-1}}{p^2}\nonumber\\
&&\times \left(\frac{-2}{nn^*}\right)^{D/2-1}\, _2F_1(1,D-3;D/2|w)
-\pi^{D/2}\Gamma(D/2-1)\nonumber\\
&&\times\frac{\Gamma(D/2-2)\Gamma(3-D/2)}{ \Gamma(D-3)} (p^2)^{D/2-3}p^-
\left(\frac{-2}{nn^*}\right)\nonumber\\
 &&\times \, _2F_1(1,D/2-1;2|w) .\end{eqnarray} 
Observe that when $D=4-2\epsilon$ there are simple poles in these two terms
but they cancel since, $$ \lim_{\epsilon\to 0}
\Gamma(\epsilon)+\Gamma(-\epsilon) = {\cal O}(1). $$

To extract the finite part one can proceed in the same way as we did in the second
paper of ref.\cite{flying}. The other representation for the integral
$B(i,j,l,m)$ also coincide with that presented in \cite{lee}. In fact, they
are related through analytic continuation (see ref.\cite{luke} for such
formulas) -- as all representations provided by NDIM for Feynman integrals in
covariant\cite{flying} and non-covariant gauges\cite{probing}.

Consider now an integral containing two factors of the form $(k\cdot n)$,

\begin{equation} T_2(i,j,l,m) = \int d^D\! q \, (q-p)^{2i}(q\cdot n)^j
\left[(q-p)\cdot n\right]^l (q\cdot n^*)^m ,\end{equation}
where the exponent $m\geq 0$. The standard approach makes use of decomposition
formulas (\ref{decomp}) and (\ref{decomp1}). On the other hand, NDIM can solve
all integrals of such kind at the same time, giving in positive
dimension\cite{prescriptionless}, 
\begin{eqnarray}
T_2(i,j,l,m) &=& \pi^{D/2} \left(\frac{2p^+ p^-}{n\cdot n^*}\right)^{D/2+i}
\frac{\Gamma(i+l+D/2) \Gamma(1+m)}{(p^+)^{-j-l}(p^-)^{-m}\Gamma(-i)
\Gamma(-j)}\nonumber\\  
&&\times \frac{\Gamma(-i-j-l-D/2)}{\Gamma(1+i+m+D/2)} ,\end{eqnarray} observe
that the  Pochhammer symbol which contain $(1+m)$ was not analytically
continued since $m$ must be either positive or zero.

\section{Conclusion}

Finally, in concluding this work we would like to emphasize some important
points concerning the application of NDIM technique to light-cone integrals.
First of all, why one needs to define the original Gaussian integral
(\ref{Gaussian}) with two-degree violation of Lorentz covariance in order to get
``causality'' preserving results? The answer is related to the very definition
of light-cone gauge condition where one has the external four-vector $n_\mu$
such that it is chosen to be light-like, i.e., $n^2=0$. However, the
``light-likeness'' condition $n^2=0$ does not uniquely define the needed
external vector $n_\mu$ to implement the gauge condition. The reason for this
is more easily seen considering a particular case where
$n_\mu=(n_0,\,n_3,\,0,\,0)$, in which case the condition $n^2=0$ gives as
solutions either $n_0=+n_3$, or $n_0=-n_3$ with $n_0>0$. Therefore, the
components of the ``light-like'' vector are not linearly independent; hence the
two possibilities: either $n_\mu\equiv (n_0,\,+n_3,\,0,\,0)$ or $n_\mu^*\equiv
(n_0,\,-n_3,\,0,\,0)$. These peculiar light-cone properties have been shown by
G. Leibbrandt \cite{NP} to be connected to the Newman-Penrose tetrad formalism
in the context of gravitation and cosmology, where a four-dimensional basis is
spanned entirely by null vectors. In his work, Leibbrandt demonstrated that the
two-dimensional vector sub-space $(n_0,\,n_3)$ {\em cannot} be spanned solely
by the vector $n_\mu$; it is not sufficient for this purpose because this
vector possesses linearly dependent components. A thorough discussion,
including not only the sub-space $(n_0,\,n_3)$, but the entire four-dimensional
space, is found in the reference above given.

This is the reason why our NDIM calculation done in \cite{probing} with only
one degree violation of Lorentz covariance failed to be ``causal'', reproducing
the well-known ``causality'' breaking PV result. The cure for this pathological
result could only be achieved by introducing the needed additional $n_\mu^*$
vector, so that the basis for spanning the entire four-dimensional vector space
be unique and well-defined. Therefore our calculations here always included
this needed dual vector $n_\mu^*$, so that our results are well-defined and
unambiguous.

This NDIM technique is therefore the most powerful and beautiful machinery
ever to come to front until now to handle the light-cone Feynman integrals.
All the desirable features are embedded in it and we can list them as follows:
{\em i)} No ``decomposition formula'' like (\ref{decomp}) is needed to
separate gauge dependent poles before the integral is evaluated; {\em ii)} It
preserves the general structure of the light-cone integral, namely, no factors
containing $n^*$ are introduced in the denominators; {\em iii)} No
prescription is needed to handle $(q\cdot n)^{-\alpha}=0$ type singularities
to solve the integral; {\em iv)} There are no parametric integrals to perform;
{\em v)} There is no need to split the dimensionality of space-time to work
out integrations over component sub-spaces like in \cite{leib-nyeo}; {\em vi)}
Results are obtained for arbitrary {\it negative} exponents of propagators
{\em and} gauge-dependent poles, so that special cases -- which
agree\cite{leib,bass,probing,prescriptionless} with the ones calculated using
ML prescription -- are contained in them; {\em vii)} Results are always within
the context of dimensional regularization, i.e., preserving gauge symmetry.

The power of NDIM is therefore readily apparent. If one remembers how integrals
over Grassmann variables are introduced, he/she can recall that the underlying
property employed to define them is translation invariance. It is an outstanding
thing that this sole property in the fermionic integration is able to guarantee
a prescriptionless light-cone and more important than that, a causality
preserving method of direct computation of Feynman integrals.

Of course, like in any gauge, two-loop integrals are more demanding when
compared to the one-loop case, and specially in the light-cone gauge.
Preliminary results we have show that NDIM can also handle them with more
easiness than in the usual standard approaches.

\section*{Acknowledgements}
 A.G.M.S. gratefully acknowledges FAPESP (Funda\c c\~ao de 
Amparo \`a Pesquisa do Estado de S\~ao Paulo, Brazil) for financial support.

\end{document}